\documentclass[a4paper,onecolumn,11pt]{IEEEtran}

\usepackage{enumitem}
\usepackage{multirow}
\usepackage{balance}
\usepackage{setspace}
\usepackage{amsmath}
\usepackage{amssymb}
\usepackage{graphicx}
\usepackage{xcolor}
\usepackage{latexsym}
\usepackage{cite}
\usepackage{url}
\usepackage{mathrsfs}
\usepackage{empheq}
\usepackage[normalem]{ulem}

\usepackage{epic,eepic}

\allowdisplaybreaks
\newtheorem{definition}{Definition}
\newtheorem{example}{Example}

\newtheorem{theorem}{Theorem}

\newtheorem{proposition}{Proposition}

\newcommand\code {{\cal C}}

\newcommand\Sc  {{\cal S}}
\newcommand\Sac {{\mathscr{S}}}

\newcommand\Tc  {{\cal T}}
\newcommand\Nc  {{\cal N}}
\newcommand\Uc  {{\cal U}}
\newcommand\Xc   {{\cal X}}
\newcommand\Yc  {{\cal Y}}

\definecolor{darkgreen}{rgb}{0, 0.5, 0}
\definecolor{poning}{rgb}{0.2,0.2,0.6}
\definecolor{poning2}{rgb}{0.6,0.2,0.2}
\definecolor{poning2d}{rgb}{0.3,0.1,0.1}

\newcounter{step}

\begin{document}

\begin{spacing}{1.5}

\title{\LARGE  {Decoder Ties Do Not Affect the Error Exponent of the Memoryless Binary Symmetric Channel}
}

\author{
\vspace{0.1in}

\IEEEauthorblockN{
     Ling-Hua Chang\IEEEauthorrefmark{1},
     Po-Ning~Chen\IEEEauthorrefmark{2},
     Fady Alajaji\IEEEauthorrefmark{3}
     and 
     Yunghsiang S.~Han\IEEEauthorrefmark{4}
\thanks{\IEEEauthorrefmark{1}Department of Electrical Engineering, Yuan Ze University, Taiwan, R.O.C. (iamjaung@gmail.com).}
\thanks{\IEEEauthorrefmark{2}Institute of Communications Engineering and Department of Electrical and Computer Engineering, National Chiao Tung University, Taiwan, R.O.C. (poningchen@nycu.edu.tw) .}
\thanks{\IEEEauthorrefmark{3}Department of Mathematics and Statistics,  Queen's University,  Kingston, ON,  Canada (fa@queensu.ca).}
\thanks{\IEEEauthorrefmark{4}Department of Communication Engineering, National Taipei University, Taiwan, R.O.C. (yunghsiangh@gmail.com).}
\thanks{
The work of Ling-Hua Chang is supported by the Ministry of Science and Technology, Taiwan, R.O.C.~(Grant\,No.\,MOST\,107-2218-E-155-012-MY2). 
The work of Po-Ning Chen is supported by the Ministry of Science and Technology, Taiwan, R.O.C.~(Grant\,No.\,MOST\,108-2221-E-009-024-MY2). 
The work of Fady Alajaji is supported by the Natural Sciences and Engineering Research Council of Canada.
} 
}
}

\end{spacing}
\maketitle

\begin{spacing}{1.5}
\begin{abstract} 
The generalized Poor-Verd\'u error lower bound established in \cite{conf} for multihypothesis testing is studied in the classical channel coding context. {It} is proved that for any sequence of block codes sent over the memoryless binary symmetric channel (BSC), the minimum probability of  error (under maximum likelihood decoding) has a relative deviation from the generalized bound that grows at most linearly in blocklength. This result directly implies that for arbitrary codes used over the BSC, decoder ties can only affect the subexponential behavior of the minimum probability of error. 
\end{abstract}

\begin{IEEEkeywords}
Binary symmetric channel, block codes, error probability bounds,  {maximum likelihood decoder ties}, error exponent, channel reliability function, hypothesis testing.
\vspace{-0.25in}
\end{IEEEkeywords}
\end{spacing}

\begin{spacing}{1.65}
\section{Introduction}
\pagestyle{plain}

A well-known lower bound on the minimum probability of error $P_{\text{e}}$ of multihypothesis testing is the so-called Poor-Verd\'{u} bound \cite{pv}. The bound was generalized in \cite{gpv} by tilting, via a parameter $\theta \ge 1$, the posterior hypothesis distribution, with the resulting bound noted to progressively improve with $\theta$ except for examples involving the memoryless binary erasure channel (BEC). The closed-form formula of this generalized Poor-Verd\'{u} bound, as $\theta$ tends to infinity, was recently derived in \cite{conf}. An alternative lower bound for $P_{\text{e}}$ was established by Verd\'{u} and Han in~\cite{vh13}; this bound was subsequently extended and strengthened in~\cite{vcfm}. 

In this paper, we investigate the generalized Poor-Verd\'{u} lower bound of~\cite{conf} in the classical context of the maximum-likelihood (ML) decoding error probability of block codes $\code_n$ with blocklength $n$ and size $|\code_n|=M$ sent over the memoryless binary symmetric channel (BSC) with crossover probability $0<p<1/2$. For convenience, we denote this lower bound by $b_n$ (see its expression in~\eqref{last-bound}). Specifically, for channel inputs uniformly distributed over code $\code_n$, we bound the code’s minimum probability of decoding error $a_n$ in terms\footnote{Note that  $a_n$ and $b_n$, as well as the notations introduced in Table~\ref{tab:1}, are all functions of the adopted code $\code_n$. For ease of notation, we drop their dependence on $\code_n$ throughout the paper.} of $b_n$ as follows:
\begin{equation}\label{key-ub}
{b_n\leq a_n\leq(1+c\,n)\,b_n},
\end{equation}
where $c \triangleq\frac{1-p}p$ is the channel (likelihood ratio) constant and is independent of code $\code_n$. Noting that $b_n$ can be recovered from $a_n$ by disregarding all decoder ties, which occur with probability no larger than $cn\cdot b_n$, we conclude that decoder ties only affect the subexponential behavior of the minimum error probability $a_n$ with respect to an arbitrary sequence of codes $\{\code_n\}_{n\geq 1}$. 

The related problem of exactly characterizing the channel reliability function at low rates remains a long-standing open problem; in-depth studies on this focal information-theoretic function and related problems include the classical papers~\cite{gallager65,SGB67-I,SGB67-II,mceliece77} and texts~\cite{gallager,viterbi,csiszar,blahut} and the more recent works~\cite{litsyn99,barg05,harout07,polyanskiy13,dalai13,fabregas14b,fabregas14,altug14,burnashev15,tandon19,tandon20,merhav20} (see also the references therein). In~\cite{pv}, Poor and Verd\'{u} conjectured that their original error lower bound for multihypothesis testing, which yields an upper bound on the channel coding reliability function, is tight for all rates and arbitrary channels. The conjecture was disproved in~\cite{vio}, where the bound was shown to be loose for the BEC at low rates. Furthermore, Polyanskiy showed in~\cite{polyanskiy13} that the original Poor-Verd\'{u} bound~\cite{pv} coincides with the sphere-packing error exponent bound for discrete memoryless channels (and is hence loose at low rates for this entire class of channels). 

The rest of the paper is organized as follows. The error bound $b_n$ is analyzed for the channel coding problem over the memoryless BSC in Section~\ref{sec:mainresult}. The proof of the main theorem is provided in detail in Section~\ref{sec:proof}. Finally, conclusions are drawn in Section~\ref{sec:conclusion}. 

Throughout the paper, we denote $[M]\triangleq\{1,2,\ldots,M\}$ for any positive integer $M$.

\section{Analysis of Lower Bound $b_n$ for an Arbitrary Sequence of Binary Codes $\{\code_n\}_{n\geq 1}$}\label{sec:mainresult}

Consider an arbitrary binary code $\code_n$ with blocklength $n$ to be used over the BSC with crossover probability $0<p< \frac 12$. It is shown in \cite[Eq.~(5)]{conf} that the lower bound $b_n$ to the minimum probability of decoding error $a_n$ is given by 
\begin{equation}\label{last-bound}
 b_n= P_{X^n,Y^n}\left\{(x^n,y^n)\in\Xc^n\times\Yc^n:P_{X^n|Y^n}(x^n|y^n)<\max_{u^n\in\code_n {\setminus\{x^n\}}}P_{X^n|Y^n}(u^n|y^n)\right\}. 
\end{equation}
Indeed, by recalling that the (optimal) maximum a posteriori (MAP) estimate of $x^n\in\code_n$ from observing $y^n\in\Yc^n$ at the channel output is given by 
\begin{equation}\label{eq:pv1}
e(y^n)=\arg \max_{x^n\in\code_n}P_{X^n|Y^n}(x^n|y^n),
\end{equation}
the right-hand-side (RHS) of~\eqref{last-bound} is nothing but the error probability under a ``genie'' MAP decoder that correctly resolves ties. We demonstrate that the lower bound $b_n$ in~\eqref{last-bound}, upon scaling it by the affine linear term $(1+c\,n)$, where $c=(1-p)/p$, becomes an upper bound for $a_n$, and hence is asymptotically exponentially tight with $a_n$ \big(i.e., $\limsup_{n\rightarrow\infty}\frac 1{n}\log\frac{a_{n}}{b_{n}}=0$\big) for arbitrary sequences of block codes sent over the BSC. The exponential tightness result follows directly from the following theorem, which is the main contribution of the paper.

\begin{theorem}\label{theorem}
For any sequence of codes  $\{\code_n\}_{n\geq 1}$ of blocklength $n$ and size $|\code_n|= {M}$ with $\code_n \subseteq \Xc^n\triangleq\{0,1\}^n$, let $a_n$ denote the minimum probability of decoding error for transmitting $\code_n$ over the BSC with crossover probability $0<p<1/2$, under a uniform distribution $P_{X^n}$ over $\code_n$, where $X^n$ is the $n$-tuple $(X_1,\ldots,X_n)$. Then,
\begin{equation}\label{eq:bnan}
b_n\leq a_n\leq \left(1+\frac{(1-p)}{p}n\right)b_n,
\end{equation}
where $b_n$ is given in \eqref{last-bound}.
\end{theorem}

In Theorem~\ref{theorem}, it is implicitly assumed that all $M$ codewords must be distinct. Note that if identical codewords are allowed in $\code_n$, decoder ties may become dominant in the minimum error probability $a_n$ and the key inequality \eqref{eq:bnan} in Theorem~\ref{theorem} no longer holds. Theorem~\ref{theorem} reveals that for any \textit{arbitrary} sequence of block codes $\{\code_n\}_{n\geq 1}$ used over the BSC, the relative deviation, $(a_n-b_n)/b_n$, of the minimum probability of decoding error $a_n$ from $b_n$ is \textit{at most linear} in the blocklength $n$. It is worth mentioning that this conclusion cannot be applied for the BEC for any code $\code_n$ because $b_n=0$ is always valid over the BEC. 

\medskip\noindent
\textit{Overview of the Proof of Theorem~\ref{theorem}}: Before providing the full proof of Theorem~\ref{theorem} in Section~\ref{sec:proof}, we introduce the necessary notation and highlight how we prove~\eqref{eq:bnan}.

Because the channel input distribution $P_{X^n}$ is uniform over $\code_n$, the code's minimal probability of error $a_n$ is achieved under ML decoding. For the BSC, the ML estimate based on any received $n$-tuple $y^n$ at the channel output is obtained via the Hamming distances $\{d(x^n,y^n)\}_{x^n\in\code_n, y^n\in\Yc^n}$. Define the set of output $n$-tuples $y^n$ which definitely lead to an ML decoder error when $x^n_{(i)}\in \code_n$ is transmitted as
\begin{equation}\label{eq:no-ties}
 \Nc_i\triangleq\bigg\{y^n\in\Yc^n: d(x^n_{(i)},y^n)>\min_{u^n\in\code_n\setminus\{x^n_{(i)}\}} d(u^n,y^n)\bigg\}, 
\end{equation}
and the set of output $n$-tuples $y^n$ that induce a decoder tie when transmitting $x^n_{(i)}\in \code_n$ as
\begin{equation}\label{p1-tn_a}
\Tc_i\triangleq\bigg\{y^n\in\Yc^n: d(x^n_{(i)},y^n)=\min_{u^n\in\code_n\setminus\{x^n_{(i)}\}} d(u^n,y^n)\bigg\}.
\end{equation} 
For the BSC with crossover probability $0<p< \frac 12$, we have $P_{Y^n|X^n}(y^n|x^n_{(i)})=\big(\frac{p}{1-p}\big)^{d(x_{(i)}^n,y^n)}(1-p)^n$. Thus, $d(x^n_{(i)},y^n)>\min_{u^n\in\code_n\setminus\{x^n_{(i)}\}}$ $d(u^n,y^n)$ if and only if $P_{Y^n|X^n}(y^n|x^n_{(i)})<\max_{u^n\in \code_n\setminus\{x^n_{(i)}\}}$ $P_{Y^n|X^n}(y^n|u^n)$, and therefore 
\begin{equation}\label{eq:bn-form}
b_n=\sum_{i=1}^M P_{X^n}(x^n_{(i)}) P_{Y^n|X^n}(\Nc_i|x^n_{(i)})=\frac 1{M}\sum_{i\in [M]}  P_{Y^n|X^n}(\Nc_i|x^n_{(i)}).
\end{equation}
Similarly, $P_{Y^n|X^n}(y^n|x^n_{(i)})=\big(\frac{p}{1-p}\big)^{d(x_{(i)}^n,y^n)}(1-p)^n$ implies that the probability of decoder ties, denoted by $\delta_n$, satisfies 
\begin{equation}\label{delta_n}
\delta_n=\sum_{i=1}^M P_{X^n}(x^n_{(i)})P_{Y^n|X^n}\big(\Tc_i|x^n_{(i)}\big )=\frac{1}{M}\sum_{i\in [M]}P_{Y^n|X^n}\big(\Tc_i|x^n_{(i)}\big).
\end{equation}
We thus obtain the following relationship: 
\begin{equation}\label{eq:anbndn}
b_n\leq a_n\leq b_n+\delta_n=\left(1+\frac{\delta_n}{b_n}\right)b_n.
\end{equation}
Note if $\delta_n=0$,\footnote{A straightforward example for which $\delta_n=0$ is $\code_n$ consisting of only two codewords whose Hamming distance is an odd number.} then~\eqref{eq:anbndn} is tight and \eqref{eq:bnan} holds trivially; so without loss of generality, we will assume in the proof that $\delta_n>0$. We then have that 
\begin{IEEEeqnarray}{rCl}
\frac{\delta_n}{b_n}&=&\frac{\sum_{i\in [M]}P_{Y^n|X^n}(\Tc_i|x^n_{(i)})}{\sum_{i\in [M]}P_{Y^n|X^n}(\Nc_i|x^n_{(i)})}\\
&\leq&\frac{\sum_{i\in [M]:\Tc_i\neq\emptyset}P_{Y^n|X^n}(\Tc_i|x^n_{(i)})}{\sum_{i\in [M]:\Tc_i\neq\emptyset}P_{Y^n|X^n}(\Nc_i|x^n_{(i)})}\label{tec1}\\
&\leq&\frac{\sum_{i\in [M]:\Tc_i\neq\emptyset}\Big(P_{Y^n|X^n}(\Nc_i|x^n_{(i)})
\cdot\max_{i'\in [M]:{\Tc_{i'}\neq\emptyset}} \frac{P_{Y^n|X^n}(\Tc_{i'}|x^n_{(i')})}{P_{Y^n|X^n}(\Nc_{i'}|x^n_{(i')})}\Big)}{\sum_{i\in [M]:\Tc_i\neq\emptyset}P_{Y^n|X^n}(\Nc_i|x^n_{(i)})}\\
&=&\max_{i'\in [M]:{\Tc_{i'}\neq\emptyset}} \frac{P_{Y^n|X^n}(\Tc_{i'}|x^n_{(i')})}{P_{Y^n|X^n}(\Nc_{i'}|x^n_{(i')})}, \label{tec2}
\end{IEEEeqnarray} 
where \eqref{tec1} holds because the assumption of $\delta_n>0$ guarantees the existence of at least one non-empty set $\Tc_i$ for $i\in[M]$. With \eqref{eq:anbndn} and \eqref{tec2}, the upper bound in \eqref{eq:bnan} follows by proving that  
\begin{equation}
\frac{P_{Y^n|X^n}(\Tc_i|x^n_{(i)})}{P_{Y^n|X^n}(\Nc_i|x^n_{(i)})}\leq \frac{(1-p)}{p}n \quad\text{for non-empty }\Tc_i.
\end{equation} 
To achieve this objective, we will construct a number of disjoint covers of $\Tc_i$ and also construct the same number of disjoint subsets of $\Nc_i$ such that a one-to-one correspondence between the $\Tc_i$-covers and the $\Nc_i$-subsets exists. Since $P_{Y^n|X^n}(\Tc_i|x^n_{(i)})>0$ guarantees the existence of at least one non-empty $\Tc_i$-cover, a similar derivation to \eqref{tec2} yields that $\frac{P_{Y^n|X^n}(\Tc_i|x^n_{(i)})}{P_{Y^n|X^n}(\Nc_i|x^n_{(i)})}$ is upper-bounded by the maximum ratio of the probabilities of the $\Tc_i$-cover-versus-$\Nc_i$-subset pairs. The final step (i.e., Proposition~\ref{proposition9} in Section~\ref{part2}) is to enumerate the probabilities of the $\Tc_i$-cover-versus-$\Nc_i$-subset pairs and show that it is bounded from above by $\frac{(1-p)}pn$ . The full details are given in the next section. 

\section{The Proof of Theorem \ref{theorem}}\label{sec:proof}

We divide the proof into four parts. In Section~\ref{part1}, we obtain a coarse disjoint covering of (non-empty) $\Tc_i$ and the corresponding disjoint subsets of $\Nc_i$. In Sections~\ref{part1-2} and \ref{part1-3}, we refine the covers of $\Tc_i$ just obtained by further partitioning each of them in a systematic manner, and the same number of disjoint subsets of $\Nc_i$ are also constructed. In Section \ref{part2}, we enumerate the refined covering sets of $\Tc_i$ and the corresponding subsets of $\Nc_i$, which enable us  to obtain the desired upper bound for $\delta_n/b_n$. Since we consider the memoryless BSC in this paper, we assume without loss of generality that $x^n_{(1)}$ is the all-zero codeword. We also assume for notational convenience that $i=1$ and $\Tc_1\neq\emptyset$.

For ease of reference, we first summarize in Table~\ref{tab:1} all main symbols used in the proof. We also illustrate in Fig.~\ref{fig1} all sets defined in Table~\ref{tab:1}, based on the code of Example~\ref{example1} below. 

\begin{table*}[t]
\begin{center}
\caption{Summary of all main symbols used in the proof.}\label{tab:1}
\begin{tabular}{|c|l|c|}
\hline\hline
{\bf Symbo}l&{\bf Description}&{\bf Definition}\\\hline\hline
$[M]$&\multicolumn{2}{|l|}{A shorthand for $\{1,2,\ldots,M\}$}\\\hline
$\code_n$&\multicolumn{2}{|l|}{The code $\big\{x_1^{(n)},x_2^{(n)},\ldots,x_M^{(n)}\big\}$ with $x_1^{(n)}$ being the all-zero codeword}\\\hline
$d(u^n,v^n|\Sc)$&\multicolumn{2}{|l|}{The Hamming distance between the portions of $u^n$ and $v^n$ with indices in $\Sc$}\\\hline\hline
\multicolumn{3}{|l|}{\emph{All terms below are functions of $\code_n$ (this dependence is not explicitly shown to simplify notation)}}\\\hline\hline
$\Nc_j$&The set of channel outputs $y^n$ that lead to an ML decoder error when $x^n_{(i)}$ is sent &\eqref{eq:no-ties} \\\hline
$\Tc_j$&The set of channel outputs $y^n$ that induce a decoder tie when $x^n_{(i)}$ is sent &\eqref{p1-tn_a}\\ \hline
$\Tc_{j|1}$&The set of channel outputs $y^n$ that are at equal distance from $x_{(1)}^n$ and $x_{(j)}^n$&\eqref{p2_a}\\
&and that are not included in $\Tc_{i|1}$ for $2\leq i\leq j-1$&\\
\hline
$\Nc_{j|1}$&The set of channel outputs $y^n$ that satisfy $d(x^n_{(1)},y^n)-1=d(x^n_{(j)},y^n)+1$&\eqref{p2_b}\\
&and that are not included in $\Nc_{i|1}$ for $2\leq i\leq j-1$&\\
\hline
$\Sc_j$&The set of indices for which the components of $x_{(j)}^n$ equal one&\\\hline
$\ell_j$&The size of $\Sc_j$, i.e., $|\Sc_j|$&\\\hline
$\Sc_{r;\lambda_r}$&It is equal to $\Sc_r$ if $\lambda_r=1$, and $\Sc_r^{\text{c}}$ if $\lambda_r=0$ \big(only used in \eqref{scjm} 
to define $\Sc_j^{(m)}$\big)&\\\hline
$\Sc_j^{(m)}$&The subset of $\Sc_j$ defined according to whether each index in $\Sc_j$ is in each&\eqref{scjm}\\
&of $\Sc_2$, $\ldots$, $\Sc_{j-2}$&\\\hline
$\Sac_j^{(m)}$&The union of $\Sc_j^{(1)}$, $\Sc_j^{(2)}$, $\ldots$, $\Sc_j^{(m)}$&\eqref{def1sac}\\\hline
$\ell_j^{(m)}$&The size of $\Sac_j^{(m)}$, i.e., $|\Sac_j^{(m)}|$&\\\hline
$\sigma(\cdot)$&The mapping from $\{0,1,\ldots,\ell_j-1\}$ to $[2^{j-2}]$ for partitioning $\Tc_{j|1}$ into $\ell_j$&\eqref{mappingkm}\\
&subsets $\{\Tc_{j|1}(k)\}_{0\leq k<\ell_j}$&\\\hline
$\Tc_{j|1}(k)$&The $k$th partition of $\Tc_{j|1}$ for $k=0$, $1$, $\ldots$, $\ell_j-1$&\eqref{p3_a}\\\hline
$\Nc_{j|1}(k)$&The $k$th subset of $\Nc_{j|1}$ for $k=0$, $1$, $\ldots$, $\ell_j-1$&\eqref{p3_b}\\\hline
$\Uc_{j|1}(k)$&The group of representative elements in $\Tc_{j|1}(k)$ for defining the partitions of $\Tc_{j|1}(k)$&\\\hline
$\Tc_{j|1}(u^n;k)$&The partition of $\Tc_{j|1}(k)$ associated with $u^n\in\Uc_{j|1}(k)$&\eqref{p4_a}\\\hline
$\Nc_{j|1}(u^n;k)$&The subset of $\Nc_{j|1}(k)$ associated with $u^n\in\Uc_{j|1}(k)$&\eqref{p4_b}\\\hline\hline
\end{tabular}
\end{center}
\end{table*}

\begin{figure*}[h]
\center
\includegraphics[width=0.7\textwidth]{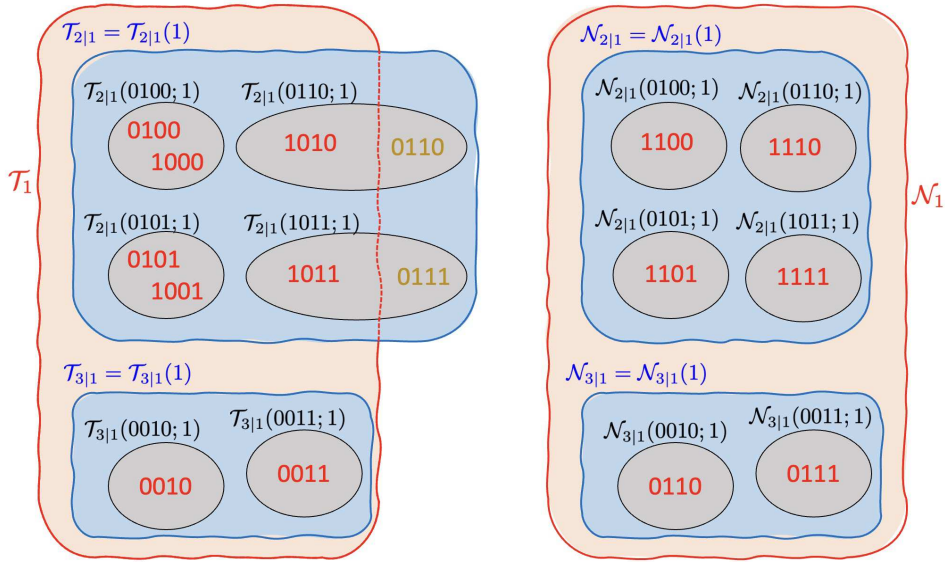}
\caption{An illustration of the sets defined in Table~\ref{tab:1}, based on the setting in Example~\ref{example1}, where $\Tc_{2|1}(0)=\Tc_{3|1}(0)=\Nc_{2|1}(0)=\Nc_{3|1}(0)=\emptyset$, $\Uc_{2|1}(1)=\{0100,0101,0110,1011\}$ and $\Uc_{3|1}(1)=\{0010,0011\}$.}\label{fig1} 
\end{figure*}

\subsection{A Coarse Disjoint Covering of Non-empty $\Tc_1$ and the Corresponding Disjoint Subsets of $\Nc_1$}\label{part1}

Before providing a coarse disjoint covering of non-empty $\Tc_1$ and corresponding disjoint subsets of $\Nc_1$, we elucidate the idea behind them.

Note from its definition in \eqref{p1-tn_a} that $\Tc_1$ consists of all minimum distance ties when $x_{(1)}^n$ is sent. To obtain disjoint covers of $\Tc_1$, we first collect all channel outputs $y^n$ that are equidistant from $x_{(1)}^n$ and $x_{(2)}^n$ and we place them in $\Tc_{2|1}$. We next place into $\Tc_{3|1}$ those outputs $y^n$ that have \emph{not} been included in $\Tc_{2|1}$, and that are at equal distance from $x_{(1)}^n$ and $x_{(3)}^n$. We iterate this process sequentially to obtain $\Tc_{j|1}$ for $j=4$, $5$, $\ldots$, $M$ by picking  $y^n$ tuples that have not yet been included in all previous collections, and that are equidistant from $x_{(1)}^n$ and $x_{(j)}^n$. This completes the construction of the disjoint covers $\{\Tc_{j|1}\}_{j=2}^M$ of $\Tc_1$. Note that for non-empty $\Tc_1$, we have at least one $\Tc_{j|1}$ that is non-empty.

The $(M-1)$ disjoint subsets of $\Nc_i$ are constructed as follows. Suppose $\Tc_{2|1}$ is non-empty. Given a channel output $u^n$ in $\Tc_{2|1}$ (that is at equal distance from $x^n_{(1)}$ and $x^n_{(2)}$), we can flip a zero component of $u^n$ to obtain a $v^n$ to fulfill $d(x^n_{(1)},v^n)-1=d(x^n_{(1)},u^n)=d(x^n_{(2)},u^n)=d(x^n_{(2)},v^n)+1$, implying $d(x^n_{(1)},v^n)>d(x^n_{(2)},v^n)\geq \min_{z^n\in \code_n\setminus\{x^n_{(1)}\}}d(z^n,v^n)$. Therefore, it follows from the definition in \eqref{eq:no-ties} that $v^n\in \Nc_1$. Collecting all such $v^n$ from every $u^n\in \Tc_{2|1}$, we form $\Nc_{2|1}$. This construction provides an operational connection between $\Tc_{2|1}$ and $\Nc_{2|1}$. Iterating this process for $j=3$, $4$, $\ldots$, $M$ in this order and deliberately avoiding repeated collections give the desired disjoint subsets of $\Nc_1$. Here, we force $\Nc_{j|1}=\emptyset$ whenever $\Tc_{j|1}$ is an empty set.  

The above constructions are formalized in the following definition. 

\begin{definition}
Define for $j\in[M]\setminus\{1\}$,
\begin{subequations}
\begin{empheq}[left=\empheqlbrace]{align}
&\Tc_{j|1}\triangleq\Big\{y^n\in\Yc^n:d(x_{(1)}^n,y^n)=d(x_{(j)}^n,y^n)
<\displaystyle\min_{r\in[j-1]\setminus\{1\}}d(x_{(r)}^n,y^n)\Big\};\label{p2_a}\\
&\Nc_{j|1}\triangleq\Big\{y^n\in\Yc^n:d(x_{(1)}^n, y^n)-1=d(x_{(j)}^n, y^n)+1\neq d(x_{(r)}^n, y^n)+1 \text{ for }r\in[j-1]\setminus\{1\}\Big\}.\label{p2_b}
\end{empheq}
\end{subequations}
\end{definition}

To better understand the terms just introduced, we provide the following example. 

\begin{example}\label{example1}
Suppose $M=3$ and $\code_4=\{x_{(1)}^4,$ $x_{(2)}^4,$ $x_{(3)}^4\}=\{0000,$ $1100,$ $0110\}$. Then, $\Tc_1=\{0100,$ $1000,$ $0101,$ $1001$, $1010$, $1011$, $0010$, $0011\}$ and $\Nc_1=\{1100,$ $0110,$ $0111,$ $1101,$ $1110,$ $1111\}$. Furthermore, we have  $\Tc_{2|1}=\{0100,$ $1000,$ $0101,$ $1001,$  $1010,$  $1011$, $0110$, $0111\}$ and $\Tc_{3|1}=\{0010,$ $0011\}$. Note that the last two elements in $\Tc_{2|1}$ satisfy both $d(x_{(1)}^n,y^n)=d(x_{(2)}^n,y^n)$ and $d(x_{(1)}^n,y^n)>d(x_{(3)}^n,y^n)$, and hence they result in \emph{ties} but not in \emph{minimum distance ties} as required for $\Tc_1$ in \eqref{p1-tn_a}, indicating that $\Tc_{2|1}\cup\Tc_{3|1}$ is a proper covering of $\Tc_1$ as shown in Fig.~\ref{fig1}. On the other hand, we have $\Nc_{2|1}=\{1100,1101,1110,1111\}$ and $\Nc_{3|1}=\{0110,0111\}$, showing that they are disjoint subsets of $\Nc_1$.\hfill$\Box$
\end{example}

The observations we made from Example~\ref{example1} are proved in the next proposition.

\begin{proposition}\label{proposition2}
For nonempty $\Tc_1$, the following two properties hold.
\begin{list}{\roman{step})}
    {\usecounter{step}
    \setlength{\labelwidth}{0.5cm}
    \setlength{\leftmargin}{1.1cm}\slshape}
\item The collection $\{\Tc_{j|1}\}_{j\in[M]\setminus\{1\}}$ forms a disjoint covering of $\Tc_{1}$.
\item $\{\Nc_{j|1}\}_{j\in[M]\setminus\{1\}}$ is a collection of disjoint subsets of $\Nc_i$.
\end{list}
\end{proposition} 
\begin{IEEEproof} 
The strict inequality in \eqref{p2_a} and the non-equality condition in \eqref{p2_b} guarantee no multiple inclusions of an element from the previous collections; therefore, $\{\Tc_{j|1}\}_{j\in[M]\setminus\{1\}}$ are disjoint and so are $\{\Nc_{j|1}\}_{j\in[M]\setminus\{1\}}$. Now for any $y^n\in\Tc_{1}$, we have $d(x_{(1)}^n,y^n)=d(x_{(m)}^n,y^n)$ for some $m\neq 1$; therefore, this $y^n$ must be collected in $\Tc_{j|1}$ for some $j\leq m$, confirming that $\{\Tc_{j|1}\}_{j\in[M]\setminus\{1\}}$ forms a covering of $\Tc_{1}$. Next, for any $y^n\in\Nc_{j|1}$, we have $d(x_{(1)}^n,y^n)-1=d(x_{(j)}^n,y^n)+1\geq\min_{u^n\in\code_n\setminus\{x_{(1)}^n\}}d(u^n,y^n)+1$, leading to $d(x_{(1)}^n,y^n)>d(x_{(j)}^n,y^n)\geq \min_{u^n\in\code_n\setminus\{x_{(1)}^n\}}d(u^n,y^n)$; hence, this $y^n$ must be contained in $\Nc_{1}$, confirming that $\{\Nc_{j|1}\}_{j\in[M]\setminus\{1\}}$ are subsets of~$\Nc_i$.
\end{IEEEproof}

From Proposition~\ref{proposition2}, we have that
\begin{equation}\label{new_43}
\frac{P_{Y^n|X^n}(\Tc_1 |x^n_{(1)})}{P_{Y^n|X^n}(\Nc_1 \Big|x^n_{(1)})}\leq \frac{ P_{Y^n|X^n}\big(\bigcup_{j\in[M]\setminus \{1\}}\Tc_{j|1}\big |x^n_{(1)}\big)}{P_{Y^n|X^n}\big(\bigcup_{j\in[M]\setminus \{1\}}\Nc_{j|1}\big  |x^n_{(1)}\big)}= \frac{\sum_{j\in[M]\setminus \{1\}} P_{Y^n|X^n}\big(\Tc_{j|1}\Big|x^n_{(1)}\big)}{\sum_{j\in[M]\setminus \{1\}} P_{Y^n|X^n}\big(\Nc_{j|i}\Big|x^n_{(1)}\big)},
\end{equation} 
which implies, using the same method to derive \eqref{tec2}, that 
\begin{equation}\label{p6}
\frac{P_{Y^n|X^n}(\Tc_{1}|x_{(1)}^n)} {P_{Y^n|X^n}(\Nc_{1} |x_{(1)}^n)} 
\leq \max_{j\in[M]\setminus\{1\}:\Tc_{j|1}\neq \emptyset}\frac{P_{Y^n|X^n}(\Tc_{j|1}|x_{(1)}^n)}{P_{Y^n|X^n}(\Nc_{j|1}|x_{(1)}^n)}\quad\text{for non-empty } \Tc_1.
\end{equation}

In the next section, we continue decomposing non-empty $\Tc_{j|1}$ and its corresponding $\Nc_{j|1}$. 

\subsection{A Partition of Non-empty $\Tc_{j|1}$ and the Corresponding Disjoint Subsets of $\Nc_{j|1}$}\label{part1-2}

For the enumeration analysis in Section~\ref{part2}, further decompositions of $\Tc_{j|1}$ and $\Nc_{j|1}$ are needed in order to facilitate the identification of which portions of $x_{(r)}^n$ are ones and which portions of $x_{(r)}^n$ are zeros for \emph{every} $r\in[j]$.  Let $\Sc_r$ denote the set of indices for which the (bit) components of $x_{(r)}^n$ equal one.

Now as an example, if we decompose $\Sc_3$ into $\Sc_2^{\text{c}}\bigcap\Sc_3$ and $\Sc_2\bigcap\Sc_3$, then we are certain that the portions of $x_{(2)}^n$ with indices in $\Sc_2^{\text{c}}\bigcap\Sc_3$
are zeros, and those with indices in $\Sc_2\bigcap\Sc_3$ are ones. Furthermore, when considering the portions of $x_{(4)}^n$ that are ones, $\Sc_4$ can be decomposed into $\Sc_2^{\text{c}}\bigcap\Sc_3^{\text{c}}\bigcap\Sc_4$, $\Sc_2^{\text{c}}\bigcap\Sc_3\bigcap\Sc_4$, $\Sc_2\bigcap\Sc_3^{\text{c}}\bigcap\Sc_4$ and $\Sc_2\bigcap\Sc_3\bigcap\Sc_4$, and the values of $x_{(2)}^n$ and $x_{(3)}^n$ are known exactly when considering their portions with indices in any of these four sets.
As such, $\Sc_4$ is partitioned into $2^{j-2}=4$ subsets (here $j=4)$. For convenience, we use the positive integer $m\triangleq 1+\sum_{r=2}^{j-1}\lambda_r\cdot 2^{r-2}$, where $1\leq m\leq 2^{j-2}$, to enumerate the $2^{j-2}$ joint intersections, where $\lambda_r=0$ implies $\Sc_r^{\text{c}}$ is involved in the joint intersections, while $\lambda_r=1$ implies $\Sc_r$ is taken instead. Thus, with $j=4$, the four sets $\Sc_2^{\text{c}}\bigcap\Sc_3^{\text{c}}\bigcap\Sc_4$, $\Sc_2\bigcap\Sc_3^{\text{c}}\bigcap\Sc_4$, $\Sc_2^{\text{c}}\bigcap\Sc_3\bigcap\Sc_4$ and $\Sc_2\bigcap\Sc_3\bigcap\Sc_4$
are respectively indexed by $m=1$, $2$, $3$ and $4$, which correspond to $(\lambda_2,\lambda_3)=(0,0)$, $(1,0)$, $(0,1)$ and $(1,1)$, respectively.  


For $j\in[M]\setminus\{1\}$, partition $\Sc_j$ into $2^{j-2}$ subsets according to whether each index in $\Sc_j$ is in $\Sc_2$, $\ldots$, $\Sc_{j-2}$ or not as follows: 
\begin{equation}\label{scjm}
\Sc_j^{(m)}\triangleq \Big(\bigcap_{r=2}^{j-2}\Sc_{r;\lambda_r}\Big)\bigcap\Sc_j \quad\text{for }1\leq m= 1+\sum_{r=2}^{j-1}\lambda_r\cdot 2^{r-2}\leq 2^{j-2},
\end{equation}
where $\Sc_{r;1}\triangleq\Sc_r$ and $\Sc_{r;0}\triangleq\Sc_r^{\text{c}}$, and each $\lambda_r\in\{0,1\}$. Define incrementally $\Sac_{j}^{(0)}\triangleq\emptyset$ and 
\begin{equation}\label{def1sac}
\Sac_j^{(m)}\triangleq\bigcup_{q=1}^m\Sc_j^{(q)}, \quad m\in[2^{j-2}].
\end{equation}
Let $\ell_j\triangleq|\Sc_j|$ and $\ell_j^{(m)}\triangleq|\Sac_j^{(m)}|$ denote the sizes of $\Sc_j$ and $\Sac_j^{(m)}$, respectively. Then, as mentioned at the beginning of this section, for all $r\in[j]$, the components of $x_{(r)}^n$ with indices in $\Sc_j^{(m)}$ can now be unambiguously identified and are all equal to $\lambda_r$. As a result, with $x_{(1)}^n$ being the all-zero codewords, 
\begin{equation}\label{remark5}
d(x_{(1)}^n,x_{(r)}^n|\Sc_j^{(m)})=\begin{cases}
|\Sc_j^{(m)}|,&\lambda_r=1;\\
0,&\lambda_r=0,
\end{cases}
\end{equation}
where $d(u^n,v^n|\Sc)$ denotes the Hamming distance between the portions of $u^n$ and $v^n$ with indices in $\Sc$, and by convention, we set $d(u^n,v^n|\Sc)=0$ when $\Sc=\emptyset$. We will see later in the proof of Proposition~\ref{proposition9} that \eqref{remark5} facilitates our evaluation of $d(x_{(r)}^n,y^n)$ for channel output $y^n$.

We illustrate the sets and quantities just introduced in the following example. 
\begin{example}\label{example2}
Suppose $\code_6=\{x_{(1)}^6,$ $x_{(2)}^6,$ $x_{(3)}^6\}=\{000000,$ $111100,$ $001111\}$. Then, from \eqref{p2_a} and \eqref{p2_b}, we obtain $\Tc_{3|1}=\{001010,$ $001001,$ $000110,$ $000101,$ $000011,$ $010011,$ $100011\}$ and $\Nc_{3|1}=\{000111$, $001011$, $001101$, $001110$, $101011$, $011011$, $100111$, $010111$, $111101$, $111110\}$. Next, it can be seen that $\Sc_{2}=\{1,2,3,4\}$, $\Sc_{3}=\{3,4,5,6\}$ and $\ell_2=\ell_3=4$. In addition, by varying $m=1+\lambda_2$ for $\lambda_2\in\{0,1\}$, $\Sc_{3}$ can be partitioned into $2^{3-1}=2$ sets, which are:
\begin{equation}
\Sc_{3}^{(m)}=\begin{cases}
\Sc_{2;0}\bigcap\Sc_{3}=\{5,6\},&m=1;\\
\Sc_{2;1}\bigcap\Sc_{3}=\{3,4\},&m=2.
\end{cases}
\end{equation}
Hence,
\begin{equation}
\Sac_3^{(m)}=\begin{cases}
\Sc_3^{(1)}=\{5,6\},&m=1;\\
\Sc_3^{(1)}\bigcup\Sc_3^{(2)}=\{3,4,5,6\},&m=2,
\end{cases}
\end{equation}
and $\ell_3^{(1)}=|\Sac_3^{(1)}|=2$ and $\ell_3^{(2)}=|\Sac_3^{(2)}|=4$. 
\hfill$\Box$
\end{example}

We are now ready to describe how we partition $\Tc_{j|1}$ and construct the corresponding disjoint subsets of $\Nc_{j|1}$. Recall from Section \ref{part1} that we can flip a zero component of $u^n$ in $\Tc_{j|1}$ to recover a $v^n$ in $\Nc_{j|1}$. This observation indicates that the number of zero components (equivalently, the number of one components) of $u^n\in\Tc_{j|1}$ with indices in $\Sac_j^{(m)}$ can be used as a factor to relate each partition of $\Tc_{j|1}$ to its corresponding subset of $\Nc_{j|1}$. As $x_{(1)}^n$ is assumed all-zero, this factor can be parameterized via $d(x_{(1)}^n,u^n|\Sac_j^{(m)})=k$ for $0\leq k<\ell_j^{(m)}$.

Irrespective of the construction of disjoint subsets of $\Nc_{3|1}$, one may improperly infer from Example \ref{example2} that $\Tc_{3|1}$ can be subdivided into $\ell_3$ partitions according to
$d(x_{(1)}^6,u^6|\Sac_3^{(1)})=k$ for each $0\leq k<\ell_3^{(1)}$,
and then according to $d(x_{(1)}^6,u^6|\Sac_3^{(1)})=\ell_3^{(1)}$ and $d(x_{(1)}^6,u^6|\Sac_3^{(2)})=k$ for $\ell_3^{(1)}\leq k<\ell_3^{(2)}=\ell_3$. However, the above setup could have two $u^6$ tuples, in respectively two different partitions of $\Tc_{3|1}$, recover the same $v^6$, leading to two {\em non-disjoint} subsets of $\Nc_{3|1}$. For example, flipping the last bit of $000110$  that belongs to the partition constrained by $d(x_{(1)}^6,000110|\Sac_3^{(1)})=1$, and flipping the 4th bit of $000011$ that is included in the partition constrained by $d(x_{(1)}^6,000011|\Sac_3^{(1)})=\ell_3^{(1)}$ and $d(x_{(1)}^6,000011|\Sac_3^{(2)})=2$ yield identical tuples given by $v^6=000111$; hence, the two partitions, indexed respectively by $k=1$ and $k=2$, recover two non-disjoint subsets of $\Nc_{3|1}$. To avoid repetitive constructions of the same $v^6$ from distinct partitions of $\Tc_{3|1}$, we note that multiple constructions of the same $v^6$ could happen only when the flipped zero component of $u^6$ is the only zero component in $\Sac_3^{(1)}$, i.e. $d(x_{(1)}^6,u^n|\Sac_3^{(1)})=\ell_3^{(1)}-1$. A solution is to place all $u^6$ tuples that result in multiple constructions of the same $v^6$ in one partition, based on which for $k\geq 2$, we refine the constraint of the $k$th partition as $\ell_3^{(1)}-1\leq d(x_{(1)}^6,u^6|\Sac_3^{(1)})\leq d(x_{(1)}^6,u^6|\Sac_3^{(2)})=k$. In this manner, $000110$ and $000011$ are both included in the partition indexed by $k=2$.

As a generalization, we constrain the $k$th partition of $\Tc_{j|1}$ by $\ell_{j}^{(m-1)}-1$ $\leq$ $d(x_{(1)}^n,u^n|\Sac_{j}^{(m-1)})$ $\leq$ $d(x_{(1)}^n,u^n|\Sac_{j}^{(m)})$ $=$ $k$ for $\ell_j^{(m-1)}-1\leq k<\ell_j^{(m)}-1$. After flipping a zero component of $u^n$ in the $k$th partition of $\Tc_{j|1}$, the resulting $v^n$ that belongs to the $k$th subset of $\Nc_{j|1}$ satisfies $\ell_{j}^{(m-1)}=d(x_{(1)}^n,v^n|\Sac_{j}^{(m-1)})\leq d(x_{(1)}^n,v^n|\Sac_{j}^{(m)})=k+1$. To simplify our set constructions in the following definition, we define the mapping from the partition index $k$ to the number $m$ satisfying $\ell_j^{(m-1)}-1\leq k<\ell_j^{(m)}-1$, which designates the set $\Sac_j^{(m)}$
the flipped zero component of $u^n$ is located in, as follows: 
\begin{equation}\label{mappingkm}
\sigma(k)\triangleq\begin{cases}
m,&\ell_j^{(m-1)}-1\leq k<\ell_j^{(m)}-1;\\
\min\big\{m:\ell_j^{(m)}=\ell_j\big\},&k=\ell_j-1.
\end{cases}
\end{equation}

\begin{definition}
Define for $k=0$, $1$, $\ldots$, $\ell_j-1$, 
\begin{subequations}
\begin{empheq}[left=\empheqlbrace]{align}
&\Tc_{j|1}(k)\triangleq
\left\{y^n\in\Tc_{j|1}: \ell_{j}^{(m-1)}-1\leq d\big(x_{(1)}^n,y^n\big|\Sac_{j}^{(m-1)}\big)\leq d\big(x_{(1)}^n,y^n\big|\Sac_{j}^{(m)}\big)=k\right\};\label{p3_a}\\
&\Nc_{j|1}(k)\triangleq
\left\{y^n\in\Nc_{j|1}: \ell_{j}^{(m-1)}=d\big(x_{(1)}^n,y^n\big|\Sac_{j}^{(m-1)}\big)\leq d\big(x_{(1)}^n,y^n\big|\Sac_{j}^{(m)}\big)=k+1\right\},\label{p3_b}
\end{empheq}
\end{subequations}
where $m=\sigma(k)$ is given in \eqref{mappingkm}.
\end{definition}

An example to illustrate the $\Tc_{j|1}$-partitions and $\Nc_{j|1}$-subsets is given below.

\begin{example}\label{example3}
Using the setting of Example~\ref{example2}, we show how we partition $\Tc_{3|1}$ according to $\Sac_{3}^{(1)}$ and $\Sac_{3}^{(2)}$ and construct the corresponding disjoint subsets of $\Nc_{3|1}$.  From \eqref{p3_a} and \eqref{p3_b}, we can obtain the partition $\{\Tc_{3|1}(k)\}_{0\leq k<\ell_3}$ and disjoint subsets $\{\Nc_{3|1}(k)\}_{0\leq k<\ell_3}$ as follows:
\begin{equation}
\Tc_{3|1}(k)=\begin{cases}
\emptyset,&k=0,1,3;\\
\Tc_{3|1},&k=2,
\end{cases}
\quad\text{and}\quad
\Nc_{3|1}(k)=\begin{cases}
\emptyset,&k=0,1,3;\\
\Nc_{3|1},&k=2,
\end{cases}
\end{equation}
as a result of the mapping
\begin{equation}
\sigma(k)=\begin{cases}
1,&\ell_3^{(0)}-1\leq k<\ell_3^{(1)}-1\ (\text{equiv. }k=0);\\
2,&\ell_3^{(1)}-1\leq k<\ell_3^{(2)}-1\ (\text{equiv. }k=1,2);\\
2,&k=\ell_3-1=3.
\end{cases}
\end{equation}
\hfill$\Box$
\end{example}

With the above definition, we next verify the partitions of non-empty $\Tc_{j|1}$ and the corresponding disjoint subsets of $\Nc_{j|1}$.

\begin{proposition}\label{proposition3}
For non-empty $\Tc_{j|1}$, the following two properties hold.
\begin{list}{\roman{step})}
    {\usecounter{step}
    \setlength{\labelwidth}{0.5cm}
    \setlength{\leftmargin}{1.1cm}\slshape}
\item $\{\Tc_{j|1}(k)\}_{0\leq k<\ell_j}$
 forms a partition of $\Tc_{j|1}$;
\item $\{\Nc_{j|1}(k)\}_{0\leq k<\ell_j}$ is a collection of disjoint subsets of $\Nc_{j|1}$.
\end{list}
\end{proposition}
\begin{IEEEproof}
It can be seen from the definitions of $\{\Tc_{j|1}(k)\}_{0\leq k<\ell_j}$ and $\{\Nc_{j|1}(k)\}_{0\leq k<\ell_j}$ that they are collections of mutually disjoint subsets of $\Tc_{j|1}$ and $\Nc_{j|1}$, respectively. It remains to show that $\Tc_{j|1}=\bigcup_{0\leq k<\ell_j}\Tc_{j|1}(k)$. Recall that $\Sac_j^{(m)}$ is a subset of $\Sc_j$ and every element $y^n$ in $\Tc_{j|1}$ must satisfy $\ell_j>d(x_{(1)}^n,y^n|\Sc_j)=d(x_{(j)}^n,y^n|\Sc_j)=\frac{\ell_j}2\geq d(x_{(1)}^n,y^n|\Sac_j^{(m)})$; hence, no element in $\Tc_{j|1}$ can fulfill $d(x_{(1)}^n,y^n|\Sac_j^{(m)})=\ell_j$. This confirms that in defining $\Tc_{j|1}(k)$ in \eqref{p3_a}, we can exclude the case of $k=\ell_j$. Since every element in $\Tc_{j|1}$ must satisfy the two constraints in $\Tc_{j|1}(k)$ for exactly one $0\leq k<\ell_j$, $\{\Tc_{j|1}(k)\}_{0\leq k<\ell_j}$ forms a partition of $\Tc_{j|1}$. 
\end{IEEEproof}

By applying a similar technique that leads to \eqref{tec2} and \eqref{p6}, Proposition~\ref{proposition3} results in the following inequality:
 
\begin{equation}\label{p7}
\frac{P_{Y^n|X^n}(\Tc_{j|1}|x_{(1)}^n)} {P_{Y^n|X^n}(\Nc_{j|1} |x_{(1)}^n)}\leq \max_{0\leq k<\ell_j:\Tc_{j|1}(k)\neq \emptyset}\frac{P_{Y^n|X^n}(\Tc_{j|1}(k)|x_{(1)}^n)}{P_{Y^n|X^n}(\Nc_{j|1}(k)|x_{(1)}^n)}\quad\text{for non-empty }\Tc_{j|1}.
\end{equation}
We further decompose non-empty $\Tc_{j|1}(k)$ and its corresponding $\Nc_{j|1}(k)$ in the next section. 

\subsection{ A Fine Partition of $\Tc_{j|1}(k)$ and the Corresponding Disjoint Subsets of $\Nc_{j|1}(k)$}\label{part1-3}

The final decomposition of $\Tc_{j|1}(k)$ and $\Nc_{j|1}(k)$ is a little involved. We elucidate its underlying concept via an example before formally presenting it. The idea is to further partition $\Tc_{j|1}(k)$ using a group of representative elements in $\Tc_{j|1}(k)$ and construct the corresponding subsets of $\Nc_{j|1}(k)$ based on the same group of representative elements. 

Pick an arbitrary element from $\Tc_{3|1}(2)$ in Example~\ref{example3} as the first representative element, say $u^6=001010$. We collect all outputs $y^6$ in $\Tc_{3|1}(2)$ such that its components with indices outside $\Sac_{3}^{(\sigma(2))}$ are exact duplications of the components of $u^6$ at the same positions, and place them in $\Tc_{3|1}(u^6;2)$. In other words, we require $d\big(u^6,y^6\big|\big(\Sac_{3}^{(2)}\big)^{\text{c}}\big)=0$.  With $\big(\Sac_3^{(2)}\big)^{\text{c}}=\{1,2\}$, we have $\Tc_{3|1}(u^6;2)=\Tc_{3|1}(001010;2)=\{000011, 001010, 001001, 000110, 000101\}$, where the first two bits of each tuple in $\Tc_{3|1}(u^6;2)$ must be equal to the first two bits of $u^6=001010$. Analogously, $\Nc_{3|1}(u^6;2)$ collects all elements in $\Nc_{3|1}(2)$ satisfying $d\big(u^6,y^6\big|\big(\Sac_{3}^{(2)}\big)^{\text{c}}\big)=0$,  and is given by $\Nc_{3|1}(001010;2)=\{000111,001011,001101,001110\}$. 
 
We can further pick another element $100011$ in $\Tc_{3|1}\setminus\Tc_{3|1}(001010;2)$ as the second representative to construct $\Tc_{3|1}(100011;2)=\{100011\}$ and the corresponding $\Nc_{3|1}(100011;2)=\{101011,100111\}$, where the first two bits of elements in the two sets must equal $10$. Continuing this process to construct $\Tc_{3|1}(010011;2)=\{010011\}$ and $\Nc_{3|1}(010011;2)=\{011011,010111\}$, we can see that all elements in $\Tc_{3|1}(2)$ have been exhausted. Thus, $\Uc_{3|1}(2)=\{001010,100011,010011\}$ is exactly the required group of representatives. 

We formalize the above set constructions in the following definition and proposition, whose proof is omitted, being a direct consequence of the construction process.

\begin{definition}\label{defi3} Define for $u^n\in\Tc_{j|1}(k)$ with $m=\sigma(k)$,
\begin{subequations}
\begin{empheq}[left=\empheqlbrace]{align}
&\Tc_{j|1}(u^n;k)\triangleq
\big\{y^n\in\Tc_{j|1}(k): d\big(u^n,y^n\Big|\big(\Sac_{j}^{(m)}\big)^{\text{c}}\big)=0\big\};\label{p4_a}\\
&\Nc_{j|1}(u^n;k)\triangleq
\big\{y^n\in\Nc_{j|1}(k): d\big(u^n,y^n\Big|(\Sac_{j}^{(m)})^{\text{c}}\big)=0\big\}.\label{p4_b}
\end{empheq}
\end{subequations}
\end{definition}

\begin{proposition}\label{proposition4} 
For non-empty $\Tc_{j|1}(k)$, there exists a group of representative $\Uc_{j|1}(k)\subseteq\Tc_{j|1}(k)$ such that the following two properties hold.
\begin{list}{\roman{step})}
    {\usecounter{step}
    \setlength{\labelwidth}{0.5cm}
    \setlength{\leftmargin}{1.1cm}\slshape}
\item $\big\{\Tc_{j|1}(u^n;k)\big\}_{u^n\in\Uc_{j|1}(k)}$  forms a (non-empty) partition of $\Tc_{j|1}(k)$;
\item $\big\{\Nc_{j|1}(u^n;k)\big\}_{u^n\in\Uc_{j|1}(k)}$  is a collection of (non-empty) disjoint subsets of $\Nc_{j|1}(k)$.
\end{list}
\end{proposition}

Again, by applying a similar technique to derive \eqref{tec2}, Proposition~\ref{proposition4} yields that for non-empty $\Tc_{j|1}(k)$, 
\begin{equation}\label{p8}
\frac{P_{Y^n|X^n}(\Tc_{j|1}(k)|x_{(1)}^n)} {P_{Y^n|X^n}(\Nc_{j|1}(k) |x_{(1)}^n)} 
\leq \max_{u^n\in\Uc_{j|1}(k)}\frac{P_{Y^n|X^n}(\Tc_{j|1}(u^n;k)|x_{(1)}^n)}{P_{Y^n|X^n}(\Nc_{j|1}(u^n;k)|x_{(1)}^n)}.
\end{equation}
What remains to confirm is that $\frac{(1-p)}pn$ is an upper bound on $\frac{P_{Y^n|X^n}(\Tc_{j|1}(u^n;k)|x_{(1)}^n)}{P_{Y^n|X^n}(\Nc_{j|1}(u^n;k)|x_{(1)}^n)}$; this will be shown in the next section.

\subsection{Characterization of a Linear Upper Bound for $\delta_n/b_n$}\label{part2}

The constraints of $\Tc_{j|1}(u^n;k)$ in \eqref{p3_a} and $\Nc_{j|1}(u^n;k)$ in \eqref{p3_b} indicate that when dealing with $\frac{P_{Y^n|X^n}(\Tc_{j|1}(u^n;k)|x_{(1)}^n)}{P_{Y^n|X^n}(\Nc_{j|1}(u^n;k)|x_{(1)}^n)}$, we only need to consider those bits with indices in $\Sac_j^{(m)}$ with $m=\sigma(k)$ because the remaining bits of all tuples in $\Tc_{j|1}(u^n;k)$ and $\Nc_{j|1}(u^n;k)$ have identical values as $u^n$. Since elements in $\Tc_{j|1}(u^n;k)$ with indices in $\Sac_j^{(\sigma(k))}$ have exactly $k$ ones, and those in $\Nc_{j|1}(u^n;k)$ with indices in $\Sac_j^{(\sigma(k))}$ have exactly $k+1$ ones, we can immediately infer that
\begin{equation}\label{eq:exact}
\frac{P_{Y^n|X^n}(\Tc_{j|1}(u^n;k)|x_{(1)}^n)}{P_{Y^n|X^n}(\Nc_{j|1}(u^n;k)|x_{(1)}^n)}=\frac{(1-p)}p\cdot\frac{|\Tc_{j|1}(u^n;k)|}{|\Nc_{j|1}(u^n;k)|}.
\end{equation}
The desired upper bound can thus be established by proving that $\frac{|\Tc_{j|1}(u^n;k)|}{|\Nc_{j|1}(u^n;k)|}\leq n$, as shown in the next proposition. 
\begin{proposition} \label{proposition9}
For non-empty $\Tc_{j|1}(u^n;k)$, we have
\begin{equation}
\frac{P_{Y^n|X^n}(\Tc_{j|1}(u^n;k)|x_{(1)}^n)}{P_{Y^n|X^n}(\Nc_{j|1}u^n;k)|x_{(1)}^n)}\leq\frac{(1-p)}pn.
\end{equation}
\end{proposition}
\begin{IEEEproof}
Recall from \eqref{p2_a}, \eqref{p3_a} and \eqref{p4_a}
that $y^n\in\Tc_{j|1}(u^n;k)$ with $m=\sigma(k)$ if and only if
\begin{subequations}
\begin{empheq}[left=\empheqlbrace]{align}
&d(x_{(1)}^n,y^n)=d(x_{(j)}^n,y^n);\label{p9-con1a_c2}\\
&d(x_{(1)}^n,y^n)<\displaystyle\min_{r\in [j-1]\setminus\{1\} }d(x_{(r)}^n,y^n);\label{p9-con1a-1_c2}\\
&\ell_{j}^{(m-1)}-1\leq d\big(x_{(1)}^n,y^n\big|\Sac_{j}^{(m-1)}\big)\leq d\big(x_{({1})}^n,y^n\big|\Sac_{j}^{(m)}\big)=k;\label{p9-con1b_c2}\\
&d\big(u^n,y^n\big|(\Sac_{j}^{(m)})^{\text{c}}\big)=0.\label{p9-con1c_c2}
\end{empheq}
\end{subequations}
Thus, we can enumerate the number of elements in $\Tc_{j|1}(u^n;k)$ by counting the number of channel outputs $y^n$ fulfilling  the above four conditions.

We then examine the number of $y^n$ satisfying \eqref{p9-con1b_c2} and \eqref{p9-con1c_c2}. Nothing that these $y^n$ have either $\ell_j^{(m-1)}-1$ ones or $\ell_j^{(m-1)}$ ones with indices in $\Sac_j^{(m-1)}$, we know there are
\begin{IEEEeqnarray}{rCl}
{\ell_{j}^{(m-1)}\choose \ell_{j}^{(m-1)}-1} {\ell_{j}^{(m)}-\ell_{j}^{(m-1)}\choose k-(\ell_{j}^{(m-1)}-1)}
+{\ell_{j}^{(m-1)}\choose \ell_{j}^{(m-1)}} {\ell_{j}^{(m)}-\ell_{j}^{(m-1)}\choose k-\ell_{j}^{(m-1)}}\label{86}
\end{IEEEeqnarray} 
of $y^n$ tuples satisfying \eqref{p9-con1b_c2} and \eqref{p9-con1c_c2}.\footnote{  To unify the expression, when $m=1$, in which case $\ell_{j}^{(0)}=0$, we assign $\binom{0}{-1}=0$
and $\binom{0}{0}=1$ in \eqref{86}. Similarly, when $k=\ell_{j}^{(m-1)}-1$, we set ${\ell_{j}^{(m)}-\ell_{j}^{(m-1)}\choose k-\ell_{j}^{(m-1)}}={\ell_{j}^{(m)}-\ell_{j}^{(m-1)}\choose{-1}}=0$.}  Considering the additional two conditions in \eqref{p9-con1a_c2} and \eqref{p9-con1a-1_c2}, we get that the number of elements in $\Tc_{j|1}(u^n;k)$ is upper-bounded by 
\eqref{86}.

On the other hand, from \eqref{p2_b}, \eqref{p3_b}, \eqref{p4_b} and $\Nc_{j|1}(u^n;k)\subseteq\Nc_{j|1}(k)\subseteq\Nc_{j|1}$, we obtain that 
$w^n\in\Nc_{j|1}(u^n;k)$ if and only if
\begin{subequations}
\begin{empheq}[left=\empheqlbrace]{align}
&d(x_{(1)}^n,w^n)-1=d(x_{(j)}^n,w^n)+1;\label{p9-con2a_c2}\\
&d(x_{(1)}^n,w^n)-1\neq d(x_{(r)}^n,w^n)+1 \text{ for }r\in[j-1]\setminus\{1\};\label{p9-con2a-1_c2}\\
&\ell_{j}^{(m-1)}=d\big(x_{(1)}^n,w^n\big|\Sac_{j}^{(m-1)}\big)\leq d\big(x_{(1)}^n,w^n\big|\Sac_{j}^{(m)}\big)=k+1;\label{p9-con2b_c2}\\
&d\big(u^n,w^n\big|(\Sac_{j}^{(m)})^{\text{c}}\big)=0.\label{p9-con2c_c2}
\end{empheq}
\end{subequations}
We then claim that any $w^n$ satisfying \eqref{p9-con2b_c2} and \eqref{p9-con2c_c2} should automatically validate \eqref{p9-con2a_c2} and \eqref{p9-con2a-1_c2}.
Note that the validity of the claim, which we prove  in Appendix \ref{detail}, immediately implies that the number of elements in $\Nc_{j|1}(u^n;k)$ can be determined by \eqref{p9-con2b_c2} and \eqref{p9-con2c_c2}, and hence 
\begin{equation}\label{36}
|\Nc_{j|1}(u^n;k)|=\binom{\ell_{j}^{(m)}-\ell_{j}^{(m-1)}}{k+1-\ell_{j}^{(m-1)}}.
\end{equation}
Under this claim, we complete the proof of the proposition using \eqref{eq:exact}, \eqref{86} and \eqref{36} as follows:
\begin{IEEEeqnarray}{rCl}
\frac{P_{Y^n|X^n}\big(\Tc_{j|1}(u^n;k)|x_{(1)}^n\big)}{P_{Y^n|X^n}\big(\Nc_{j|1}(u^n;k)|x_{(1)}^n\big)}
&\leq&\frac{(1-p)}p\cdot\frac{{\ell_{j}^{(m-1)}\choose \ell_{j}^{(m-1)}-1} {\ell_{j}^{(m)}-\ell_{j}^{(m-1)}\choose k-(\ell_{j}^{(m-1)}-1)}
+{\ell_{j}^{(m-1)}\choose \ell_{j}^{(m-1)}} {\ell_{j}^{(m)}-\ell_{j}^{(m-1)}\choose k-\ell_{j}^{(m-1)}}}{\binom{\ell_{j}^{(m)}-\ell_{j}^{(m-1)}}{k+1-\ell_{j}^{(m-1)}}}\label{morezero}\\
&=&\frac{(1-p)}p\bigg (\ell_{j}^{(m-1)}+\frac{k+1-\ell_{j}^{(m-1)}}{\ell_{j}^{(m)}-k}\bigg )\\
&\leq&\frac{(1-p)}p\Bigg( \ell_{j}^{(m-1)}+\frac{\ell_{j}^{(m)}-\ell_{j}^{(m-1)}}1\Bigg)\label{so_1}\\
&\leq&\frac{(1-p)}p\,n\label{so_2},
\end{IEEEeqnarray}
where \eqref{so_1} holds because $\ell_{j}^{(m-1)}-1\leq k\leq\ell_{j}^{(m)}-1$ by \eqref{mappingkm}, and \eqref{so_2} follows from $\ell_j^{(m)}\leq \ell_j\leq n$. 
\end{IEEEproof}

Using \eqref{p6}, \eqref{p7}, \eqref{p8} and Proposition~\ref{proposition9}, we obtain
\begin{equation}\label{eq:final}
\frac{P_{Y^n|X^n}(\Tc_1|x_{(1)}^n)}{P_{Y^n|X^n}(\Nc_1|x_{(1)}^n)}\leq\frac{(1-p)}pn.
\end{equation} \color{black}
We close this section by remarking that the same inequality as \eqref{eq:final}, i.e., 
\begin{equation}
\frac{P_{Y^n|X^n}(\Tc_i|x_{(i)}^n)}{P_{Y^n|X^n}(\Nc_i|x_{(i)}^n)}\leq\frac{(1-p)}pn,
\end{equation} \color{black}
can be analogously established for all $i\in[M]$ with $\Tc_i\neq\emptyset$.
Consequently, \eqref{tec2} implies
\begin{equation}
\frac{\delta_n}{b_n}
\leq\max_{i\in[M]: \Tc_i\neq\emptyset} \frac{P_{Y^n|X^n}(\Tc_i|x_{(i)}^n)}{P_{Y^n|X^n}(\Nc_i|x_{(i)}^n)}\leq \frac{(1-p)}pn.
\end{equation}

\section{Conclusion}
\label{sec:conclusion}
In this paper, the generalized Poor-Verd\'u error lower bound of \cite{conf} was considered in the classical channel coding context over the BSC. We proved that the bound is exponentially tight in blocklength as a direct consequence of a key inequality, showing that for any block code used over the BSC, the relative deviation of the code's minimum probability of error from the lower bound grows at most linearly in blocklength. 

Even though the exact determination of the reliability function of the BSC at low rates remains a daunting open problem, 
our results offer potentially a new perspective or tool for subsequent studies.
Other future work includes investigating 
sharp bounds for codes with small-to-moderate
blocklengths (e.g., see~\cite{polyanskiy,CLM13,vcfm}) used over  {symmetric} channels.  As our counting analysis for the binary symmetric channel relies heavily on the equivalence between ML decoding and minimum Hamming distance decoding, which does not hold for non-symmetric channels, extending our results to general channels may require more sophisticated enumerating techniques.
  
  
\begin{appendices}
\section{The Proof of \eqref{p9-con2b_c2} and \eqref{p9-con2c_c2} 
implying \eqref{p9-con2a_c2} and \eqref{p9-con2a-1_c2}}
\label{detail}

We validate the claim via the construction of an auxiliary $v^n\in \Nc_{j|1}(u^n;k)$ from $u^n\in \Tc_{j|1}(u^n;k)$. This auxiliary $v^n$ will be defined differently according to whether $d\big(x_{(1)}^n,u^n\big|\Sac_{j}^{(m-1)}\big)$ equals $\ell_{j}^{(m-1)}$ or $\ell_{j}^{(m-1)}-1$ as follows.
\begin{enumerate}
\item[$i)$] $d(x^n_{(1)},u^n|\Sac_{j}^{(m-1)})=\ell_{j}^{(m-1)}$: Since in this case, $u^n$ has no zero components with indices in $\Sac_j^{(m-1)}$, we flip a zero component of $u^n$ with its index in $\Sac_{j}^{(m)}\setminus \Sac_{j}^{(m-1)}=\Sc_{j}^{(m)}$ to construct a $v^n$ such that  
\begin{equation}\label{key-extension}
d(x_{(1)}^n,v^n)=d(x_{(1)}^n,u^n)+1\quad\text{and}\quad
d(x_{(j)}^n,v^n)=d(x_{(j)}^n,u^n)-1,
\end{equation}
where the existence of such $v^n$ is guaranteed by $k\leq\ell_{j}^{(m)}-1$. Then, $v^n$ must fulfill \eqref{p9-con2a_c2}, \eqref{p9-con2b_c2} and \eqref{p9-con2c_c2} (with $w^n$ replaced by $v^n$) as $u^n$ satisfies \eqref{p9-con1a_c2}, \eqref{p9-con1b_c2} and \eqref{p9-con1c_c2}. We next prove $v^n$ also fulfills \eqref{p9-con2a-1_c2} by contradiction. Suppose there exists a $r\in[j-1]\setminus\{1\}$ satisfying 
\begin{equation}\label{contradiction}
d(x_{(1)}^n,v^n)-1=d(x_{(r)}^n,v^n)+1.
\end{equation}
We then recall from \eqref{remark5} that $d(x_{(1)}^n,x_{(r)}^n|\Sc_j^{(m)})$ is either $0$ or $|\Sc_j^{(m)}|$. Thus, \eqref{contradiction} can be disproved by differentiating two cases: $1)$ $d(x_{(1)}^n,x_{(r)}^n|\Sc_j^{(m)})=0$, and $2)$ $d(x_{(1)}^n,x_{(r)}^n|\Sc_j^{(m)})=|\Sc_j^{(m)}|$.

\hspace*{6mm}
In case $1)$,  $v^n$ that is obtained by flipping a zero component of $u^n$ with index in $\Sc_j^{(m)}$ must satisfy $d(x_{(1)}^n,v^n)=d(x_{(1)}^n,u^n)+1$ and $d(x_{(r)}^n,v^n)=d(x_{(r)}^n,u^n)+1$. Then, \eqref{contradiction} implies $d(x_{(1)}^n,u^n)-1=d(x_{(r)}^n,u^n)+1$. A contradiction to the fact that $u^n$ satisfies \eqref{p9-con1a-1_c2} is obtained. In case $2)$, the flipping manipulation on $u^n$ results in $d(x_{(1)}^n,v^n)=d(x_{(1)}^n,u^n)+1$ and $d(x_{(r)}^n,v^n)=d(x_{(r)}^n,u^n)-1$. Therefore, \eqref{contradiction} implies $d(x_{(1)}^n,u^n)=d(x_{(r)}^n,u^n)$, which again contradicts \eqref{p9-con1a-1_c2}. Accordingly, $v^n$ must also fulfill \eqref{p9-con2a-1_c2}; hence, $v^n\in\Nc_{j|1}(u^n;k)$.

\hspace*{6mm}
With this auxiliary $v^n$, we are ready to prove that every $w^n$ satisfying \eqref{p9-con2b_c2} and \eqref{p9-con2c_c2} also validates \eqref{p9-con2a_c2} and \eqref{p9-con2a-1_c2}. This can be done by showing $d(x_{(r)}^n,w^n)=d(x_{(r)}^n,v^n)$ for every $r\in [j]$, which can be verified as follows:
\begin{IEEEeqnarray}{rCl}
d(x_{(r)}^n,w^n)
&=&d\big(x_{(r)}^n,w^n\big|\Sac_{j}^{(m-1)}\big)+d\big(x_{(r)}^n,w^n\big|\Sc_{j}^{(m)}\big)+d\big(x_{(r)}^n,w^n\big|(\Sac_{j}^{(m)})^{\text{c}}\big)\label{pre-con1}\\
&=&d\big(x_{(r)}^n,v^n\big|\Sac_{j}^{(m-1)}\big)+d\big(x_{(r)}^n,v^n\big|\Sc_{j}^{(m)}\big)+d\big(x_{(r)}^n,v^n\big|(\Sac_{j}^{(m)})^{\text{c}}\big)\label{con1}\\
&=&d(x_{(r)}^n,v^n)\label{after-con1},
\end{IEEEeqnarray}
where the substitution in the first term of \eqref{con1} holds because both $v^n$ and $w^n$ satisfy \eqref{p9-con2b_c2}, implying all components of $v^n$ and $w^n$ with indices in $\Sac_j^{(m-1)}$ are equal  to one; the substitution in the 2nd term of \eqref{con1} holds because when considering only those portions with indices in $\Sc_j^{(m)}$, $x_{(r)}^n$ are either all ones or all zeros according to \eqref{remark5}, and both $w^n$ and $v^n$ have exactly $k+1-\ell_j^{(m-1)}$ ones according to \eqref{p9-con2b_c2}; and the substitution in the 3rd term of \eqref{con1} is valid since both $v^n$ and $w^n$ satisfy \eqref{p9-con2c_c2}. 

\item[$ii)$] $d(x^n_{(1)},u^n|\Sac_{j}^{(m-1)})=\ell_{j}^{(m-1)}-1$: Now we let $v^n$ be equal to $u^n$ in all positions but one in $\Sac_{j}^{(m-1)}$ such that $d(x^n_{(1)},v^n|\Sac_{j}^{(m-1)})=\ell_{j}^{(m-1)}$. Then, $v^n$ must fulfill \eqref{p9-con2a_c2}, \eqref{p9-con2b_c2} and \eqref{p9-con2c_c2} as $u^n$ satisfies \eqref{p9-con1a_c2}, \eqref{p9-con1b_c2} and \eqref{p9-con1c_c2}. With the components of $x_{(r)}^n$ with respect to $\Sc_{j}^{(m)}$ being either all zeros or all ones, the same contradiction argument after \eqref{contradiction} can disprove the validity of \eqref{contradiction} for this $v^n$ and for any $r\in [j-1]\setminus\{1\}$. Therefore, $v^n$ also fulfills \eqref{p9-con2a-1_c2}, implying $v^n\in\Nc_{j|1}(u^n;k)$. With this auxiliary $v^n$, we can again verify \eqref{after-con1} via the same derivation in \eqref{after-con1}. The claim that 
$w^n$ satisfying \eqref{p9-con2b_c2} and \eqref{p9-con2c_c2} validates \eqref{p9-con2a_c2} and \eqref{p9-con2a-1_c2} is thus confirmed.
\end{enumerate}
\end{appendices}

\bibliographystyle{IEEEtran}
\bibliography{biblio_dis_pn}
\end{spacing}
\end{document}